# BEM-based SMS-LORETA –
# an advanced method to localize multiple simultaneously active sources in the cerebral cortex

Avni Pllana and Herbert Bauer[*]

Biological Psychology Unit,

Department of Clinical, Biological and Differential Psychology,

University of Vienna

Liebiggasse 5, 1010 Vienna, Austria.

Running head: BEM-based SMS-LORETA Evaluation

[*]Corresponding author,  E-mail: herbert.bauer@univie.ac.at, Fax: ++4314277-47939

# BEM-based SMS-LORETA –

## an advanced method to localize multiple simultaneously active sources in the cerebral cortex

Avni Pllana and Herbert Bauer

**Abstract**:

In this paper the method and performance data of 'Boundary Element Method (BEM)'-based SMS-LORETA (Simultaneous Multiple Sources LORETA) are presented. According to these data the method is capable of locating efficiently multiple simultaneously active neural sources from scalp potential topographies automatically. BEM-based SMS-LORETA is a procedure to fully interpret sLORETA solutions, i.e., with a given scalp potential distribution it gives the number of identifiable sources as well as their strength and orientation. Performance data result from numerous analyses of simulated noise-free and noise-contaminated potential distributions (topographies) that have been obtained by means of BEM-based forward solutions, where one, two or three simultaneously active dipoles were randomly chosen regarding their positions and polarity.

**Introduction**

BEM-based SMS-LORETA (principles in Pllana and Bauer, 2008) – a sLORETA version (Pascual-Marqui, 2002) – was developed as an effort to enable EEG-based real Local Brain Activity (LBA-) feedback training. In principle, LBA-feedback consists of multi-channel EEG recording at a particular sampling rate and its online translation into its 3D generator pattern after preprocessing, e.g. eye movement artifact rejection, low pass filtering, etc. Only if a sufficiently strong generator (source) appears within a predefined region of interest (ROI) trainees receive some sensory feedback, e.g. a gray square on a monitor turns green or a tone gets louder. At the beginning of that kind of neurofeedback training, progress is fully dependent on the spontaneous activity rate in the ROI. With a realistic ROI size, e.g. BA (Brodmann Area) 24 and 32, the estimated probability of spontaneous sources localized within the ROI is quite low. Since the feedback-stimulus additionally causes local activity outside the ROI, it is indispensable to identify preferably all sources that contribute to a particular scalp potential topography not just the most prominent. And indeed, preliminary observations in our LBA-feedback experiments so far show that two generator patterns appear quite frequently and even triple patterns occur.

The capability of a source localization method to identify multiple simultaneously active sources is of general interest, and BEM-based SMS-LORETA commends itself as a generally applicable analysis tool. For example, brain activities frequently occur simultaneously in homologous regions of both hemispheres. Since these activities are mostly unequal in strength in both hemispheres due to the functional lateralization bias and/or the asymmetry of homologous regions in terms of anatomy, gyrification and morphology they usually induce topographies that carry sufficient information to enable separation of such pairs of simultaneously active sources. BEM-based SMS-LORETA is efficient in this respect, too.

**Methods**

For all tests performed in this simulation study the SMS-LORETA algorithm (Pllana & Bauer, 2008) was implemented in MATLAB integrating the boundary element method (BEM) (Hämäläinen and Sarvas, 1989). A realistic head model was derived from averaged T1-weighted magnetic resonance (MR) head images, provided by the Montreal Neurological Institute (MNI), available in FSL (http://www.fmrib.ox.ac.uk/fsl/) standard data folder (MNI152_T1_2mm.nii.gz).

The head model is composed of three compartments: the scalp, the skull and the brain surrounded by the meninges and the subarachnoidal space. For computing the forward gain matrix the following conductivity values have been used: $\sigma = 0.33 (\Omega m)^{-1}$ for brain, meninges, subarachnoidal space and scalp conductivity, and $\sigma_{sk} = 0.0042\,(\Omega m)^{-1}$ for skull conductivity. The electrode locations were obtained by radial projections of EASYCAP (see References) montage #10 spherical coordinates from the center of the head to the scalp as indicated in Fig. 1 by black dots. The solution space is represented by a regular cubic grid of points or voxels with a spacing of 8 mm, restricted to the gyrified part of the cerebral cortex – see Fig. 1 and 2A. The number of grid points was 2712 and the number of sensors was 59.

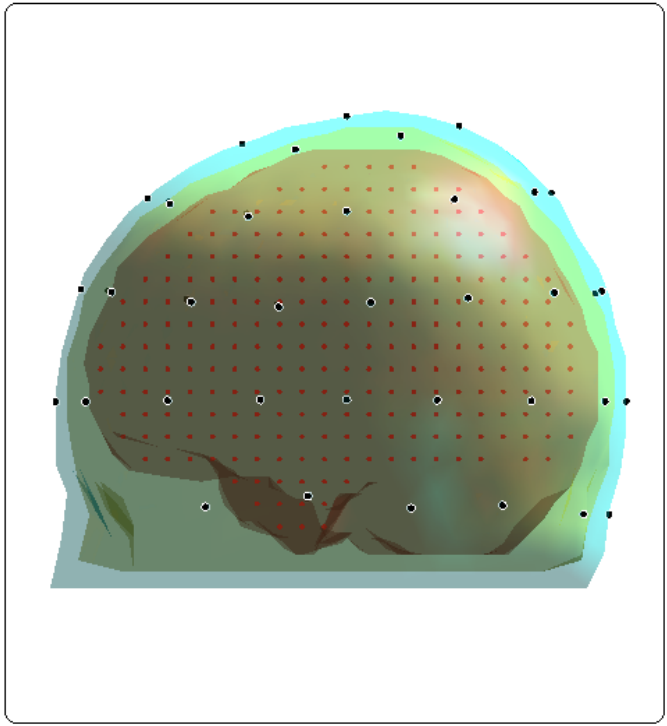

Fig. 1: 3D image of the head model showing scalp (blue), skull (olive) and brain plus meninges and subarachnoidal space (brown) in a sagittal view; black dots indicate electrode locations; red dots mark cubic grid points of the solution space.

The following pseudo code concisely illustrates the iterative SMS-LORETA algorithm (Pllana & Bauer, 2008):

$\mathbf{F} = \mathbf{\Phi}$                         /∗ recorded/simulated scalp potential distribution

$k = 0$

$\hat{\mathbf{J}} = \mathbf{TF}$                      /∗ first step of initial sLORETA

$\hat{\mathbf{J}} = \mathbf{S}_d\hat{\mathbf{J}}$                   /∗ second step of initial sLORETA

$\left|\hat{\mathbf{J}}_{i=i_0}\right| = \max_{i=1,\ldots,M}\left|\hat{\mathbf{J}}_i\right|$        /∗ tagging maximum strength voxel in initial sLORETA solution

$m = i_0$

while $\left| \mathbf{F} \right| > 0.05 \left| \boldsymbol{\Phi} \right|$

      $k = k + 1$

      $\mathbf{F}_b = \mathbf{K}_m \hat{\mathbf{J}}_m$           /* forward solution of maximum strength voxel value $\hat{\mathbf{J}}_m$

      $\mathbf{F} = \mathbf{F} - \mathbf{F}_b$           /* negative feedback - subtract forward solution

      $\hat{\mathbf{J}} = \mathbf{T}\mathbf{F}$           /* first step of sLORETA

      $\hat{\mathbf{J}} = \mathbf{S}_d \hat{\mathbf{J}}$           /* second step of sLORETA

      $\left| \hat{\mathbf{J}}_{i=i_k} \right| = \max\limits_{i=1,\dots,M} \left| \hat{\mathbf{J}}_i \right|$       /* tagging maximum strength voxel

      $m = i_k$

      $\mathbf{I}(k) = i_k$           /* store index of this voxel

end

$\mathbf{I}$ denotes the vector of indices of locations that occurred as maximum strength voxel during this iterative process. Sorting the indices in $\mathbf{I}$ according to the frequency of their occurrence in descending order gives vector $\mathbf{I}_s$. From $\mathbf{I}_s$ we obtain the matrix $\mathbf{L}_n = \mathbf{L}(\mathbf{I}_s(1 \cdots n))$ of locations of the $n$ as strongest estimated current dipoles, where $\mathbf{L} \in R^{M \times 3}$ is the matrix of Cartesian coordinates of grid points in the solution space.

## Evaluation

### *Simulation conditions*

BEM-based SMS-LORETA was evaluated in three experiments by 3 runs each. Each run consisted of a series of 1000 simulated 'scalp' potential topographies generated by 1, 2 or 3 (see Fig. 2A) simultaneously active dipoles, respectively. With all these simulated data the spatial orientation of the sources was radial but their polarity and position within the solution space was chosen fully randomly. In order to achieve more realistic, EEG/ERP-like simulations, we used measurement noise and the strength of the sources, i.e., their [dipole] moments $\vec{j}$ were a function of their voxel (source) position $\vec{r}$ according to the following relation:

$$\vec{j} = c \cdot \vec{r} \cdot \left( \frac{40}{\left| \vec{r} \right|} \right)^2 \cdot \text{sign}(\text{rand} - 0.5).$$

This relation led to stronger sources with increasing radial depth and kept their contribution to the simulated 'scalp' potential topographies comparable. In the head model the voxel (source) position radius $|\vec{r}|$ varies between 9 and 45, so we have chosen $|\vec{r}| = 40$ as reference radius. Since the measurement noise has been normalized to a root mean square ($rms$) value of 1, c is a constant multiplier in order to obtain a $SNR \approx 10$ with one source.

$SNRs$ were calculated as the ratio of the $rms$ value of the potential amplitudes at all electrode locations and the $rms$ value of the noise added at each of these locations for a given sampling point:

$$SNR = \frac{rms(Signal)}{rms(Noise)} \quad .$$

### Simulation experiments

In order to demonstrate the limits of BEM-based SMS-LORETA noise-free simulated 'scalp' potential topographies were evaluated in the first experiment. In the other two experiments – one using 59 and the other using 36 electrodes – averaged spontaneous multi-channel EEG signals were added as noise to achieve realistic, 'ERP-like' topography simulations as shown in Fig. 2B.

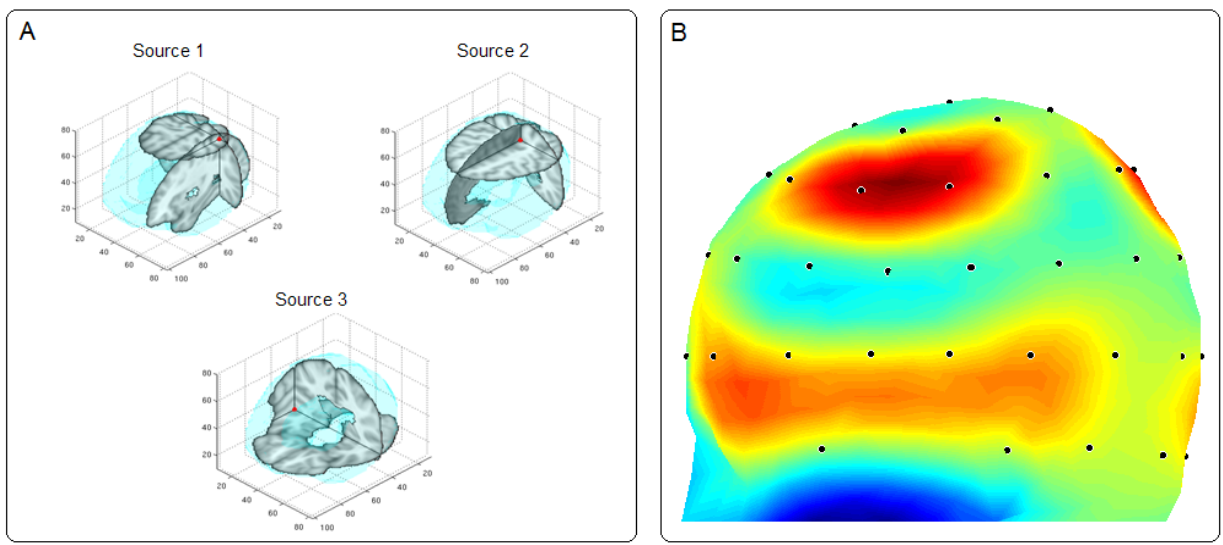

Fig. 2: A: Visualization of the locations of a set of three generating dipoles as intersections of three orthogonal brain slices; see also the solution space indicated by the brain slices.
B: Example of a potential distribution generated by BEM-based forward solution using 3 dipoles; black dots indicate electrode locations.

## Results

The statistical results of the first experiment with 1, 2 and 3 generating dipoles using 59 electrodes are summarized on Table 1. The regularization parameter was set to $\alpha = 0.0$ in all three runs since no noise was added.

**Table 1**

Zero noise, $\alpha = 0$.

1 generating source

|  | Localization performance with maximum error of 1 voxel in [%] |
|---|---|
| 1 source | 100 |

2 generating sources

|  | Localization performance with maximum error of 1 voxel in [%] |
|---|---|
| 2 sources | 99 |
| at least 1 of 2 sources | 100 |

3 generating sources

|  | Localization performance with maximum error of 1 voxel in [%] |
|---|---|
| 3 sources | 88 |
| at least 2 of 3 sources | 99 |
| at least 1 of 3 sources | 100 |

Table 1: Localization performance for a series of 1000 simulated noise free scalp potential distributions generated by 1 dipole and 2 and 3 simultaneously active dipoles using 59 recording electrodes.

The results of experiment 2 utilizing 59 electrodes are summarized in Table 2. In this case the regularization parameter was set to $\alpha = 0.001$, 0.00001 and 0.000001 with 1, 2 and 3 generating sources, respectively. The optimal values for $\alpha$ have been determined by simulation.

**Table 2**

59 Electrodes, averaged spontaneous EEG added as noise

1 generating source, α = 0.001

| $SNR_{mean}$ = 9.65<br>$SNR_{stdev}$ = 2.62 | Localization performance<br>with maximum error of 1 voxel<br>in [%] |
|---|---|
| 1 source | 99 |

2 generating sources, α = 0.00001

| $SNR_{mean}$ = 13.48<br>$SNR_{stdev}$ = 3.21 | Localization performance<br>with maximum error of 1 voxel<br>in [%] |
|---|---|
| 2 sources | 67 |
| at least 1 of 2 sources | 91 |

3 generating sources, α = 0.000001

| $SNR_{mean}$ = 16.1<br>$SNR_{stdev}$ = 3.67 | Localization performance<br>with maximum error of 1 voxel<br>in [%] |
|---|---|
| 3 sources | 38 |
| at least 2 of 3 sources | 79 |
| at least 1 of 3 sources | 97 |

Table 2: Localization performanve for a series of 1000 simulated and noise contaminated scalp potential distributions generated by 1 dipole and 2 and 3 simultaneously active dipoles using 59 recording electrodes.

And the results of experiment 3 applying 36 electrodes are summarized in Table 3. For comparability the regularization parameter was set to α = 0.001, 0.00001 and 0.000001 with 1, 2 and 3 generating sources, respectively. The optimal values for α have been determined by simulation.

**Table 3**

36 Electrodes, averaged spontaneous EEG added as noise

1 generating source, α = 0.001

| $SNR_{mean}$ = 9.23<br>$SNR_{stdev}$ = 2.0 | Localization performance<br>with maximum error of 1 voxel<br>in [%] |
|---|---|
| 1 source | 98 |

2 generating sources, α = 0.00001

| $SNR_{mean}$ = 13.24<br>$SNR_{stdev}$ = 3.5 | Localization performance<br>with maximum error of 1 voxel<br>in [%] |
|---|---|
| 2 sources | 50 |
| at least 1 of 2 sources | 83 |

3 generating sources, α = 0.000001

| $SNR_{mean}$ = 15.51<br>$SNR_{stdev}$ = 3.54 | Localization performance<br>with maximum error of 1 voxel<br>in [%] |
|---|---|
| 3 sources | 20 |
| at least 2 of 3 sources | 56 |
| at least 1 of 3 sources | 94 |

Table 3: Localization performance for a series of 1000 simulated and noise contaminated scalp potential distributions generated by 1 dipole and 2 and 3 simultaneously active dipoles using 36 recording electrodes.

**Discussion**

As already mentioned, SMS-LORETA offers an efficient way to fully and automatically analyze sLORETA transformations, i.e., estimated 3D current density equivalents of scalp potential distributions. The results of Experiment 1 (see Table 1) have been included to demonstrate the limits of this method under ideal, noise-free signal conditions. Although one source at least could always be localized in all three conditions, i.e., 1, 2 and 3 generating sources, with identifying both generating sources the performance dropped already to 99% (i.e., 10 misses in 1000 applications); the localization of all three generating sources was in 88% (i.e., already 120 misses in 1000 applications) correct (within one voxel tolerance) and two of three generating sources could be correctly identified in 99% of the applications. This was the result we expected, of course – the more complex the generator configuration the more ambiguous/erroneous the inverse solution.

An even more degrading influence on the accuracy of inverse solutions has *noise* – that's a well-known fact, of course, but again demonstrated by Experiment 2. However, having neuro-feedback applications in mind, the results of Experiment 2 look quite promising. In the LBA-feedback applications, which we ran in our lab successfully by now, beside single sources, double source patterns occurred quite frequently whereas triple sources appeared more rarely – therefore the hit rate of about 70 – 80% for double sources with the  measurement noise contaminated data seems to be perfectly sufficient.

Also the number of electrodes used is an important issue  in connection with frequent and also recurrent applications within one subject, i.e., within LBA-feedback. As can be seen in the results of Experiment 3 the localization performance for  double source patterns dropped to 50 – 55%, but the single source hit rate was still between 83 and 94% - if  that is sufficient to start and maintain LBA-feedback learning remains to be seen.

 As a concluding remark, however, we see BEM-based SMS-LORETA - using a sufficient number of electrodes presumed - as particularly suited for LBA-feedback and appropriate as an ERP analysis tool, too.

## Acknowledgments

This research was supported by grants of the Austrian Science Fund (FWF grant # P19830-B02) and the Austrian National Bank (Jubiläumsfonds grant # 12475).